\documentclass{article}
\usepackage{spconf,amsmath,amssymb,graphicx,bm}
\usepackage{adjustbox,algorithm, caption,algorithmic, color,soul}

\DeclareMathOperator*{\argmin}{arg\,min}
\DeclareMathOperator{\prox}{prox}
\newcommand{\defn}{\triangleq}
\usepackage{hyperref}

\title{MRI Recovery with A Self-calibrated Denoiser}
\name{Sizhuo Liu$^\ast$, Philip Schniter$^\dagger$, and Rizwan Ahmad$^\ast$ \thanks{This work was funded by the National Institute of Health grants R01HL135489 and R01EB029957.}}
\address{$^{\star}$Department of Biomedical Engineering, Ohio State University, Columbus OH, 43210, USA \\
			    $^{\dagger}$Department of Electrical and Computer Engineering, Ohio State University, Columbus OH, 43210, USA}

\begin{document}
%
\maketitle
\begin{abstract}
Plug-and-play (PnP) methods that employ application-specific denoisers have been proposed to solve inverse problems, including MRI reconstruction. However, training application-specific denoisers is not feasible for many applications due to the lack of training data. In this work, we propose a PnP-inspired recovery method that does not require data beyond the single, incomplete set of measurements. The proposed method, called recovery with a self-calibrated denoiser (ReSiDe), trains the denoiser from the patches of the image being recovered. The denoiser training and a call to the denoising subroutine are performed in each iteration of a PnP algorithm, leading to a progressive refinement of the reconstructed image. For validation, we compare ReSiDe with a compressed sensing-based method and a PnP method with BM3D denoising using single-coil MRI brain data.
\end{abstract}
\section{introduction}
\label{sec:intro}
Magnetic resonance imaging (MRI) suffers from slow data acquisition, which can lead to longer scan time and compromised spatial or temporal resolution. The data acquisition for MRI occurs in the spatial Fourier domain, called k-space, and can be accelerated by prospective undersampling. Compressed sensing (CS) has been extensively applied to facilitate image recovery from highly undersampled k-space data. All major MRI vendors now offer CS-inspired reconstruction methods on their commercial scanners. More recently, deep learning (DL)-based reconstruction methods have been shown to outperform CS in terms of image quality. Typically, these methods rely on supervised learning to train a convolutional neural network (CNN) that recovers images from undersampled k-space data or aliased images \cite{yang2017dagan}. Other supervised learning techniques are inspired by variational optimization methods where an iterative algorithm is unrolled and iterates between data-consistency enforcement and CNN application, which provides regularization \cite{hammernik2018learning}.

Despite the improvements offered by DL-based methods, their extension to applications where training data are scarce remains challenging. For example, collecting fully sampled k-space data for dynamic applications with time-varying physiology is challenging. In addition, the test data may have significantly different image contrast, resolution, or anatomical features compared to the training database. This mismatch between the training and testing data can lead to performance degradation. Plug-and-play (PnP) methods, which solve an inverse problem by calling a denoising subroutine in each iteration, can circumvent this problem. PnP methods do not require training data when an off-the-shelf denoiser, e.g., BM3D \cite{danielyan2011bm3d}, is used. However, more recently, it has been demonstrated that the performance of PnP can be improved by employing an application-specific denoiser, typically trained using DL \cite{liu2020free}. These denoisers do not require fully sampled k-space data but instead can be trained on images or image patches. However, these high-quality images or image patches may still not be available for some applications, especially dynamic applications. 

Recently, several methods have been proposed that utilize undersampled k-space data for training \cite{tamir2019unsupervised, cole2020unsupervised}. For example, using the Noise2Noise approach, Lehtinen et al. demonstrated that a denoiser or a network that performs image recovery can be trained from input and the target images that are both corrupted with k-space undersampling or noise \cite{lehtinen2018noise2noise}; however, acquiring the data twice for the same imaging slice is not always practical. A broader challenge with these methods is that they still require training data, and databases of undersampled MRI measurements are not readily available.

More recently, there has been an increased interest in developing self-supervised methods that require no training data at all \cite{jin2019time, yaman2020self}. For example, Yaman et al. have proposed a self-supervised learning method that partitions the undersampled k-space into multiple subsets and trains a network that infers one subset from the other \cite{yaman2020self}. This method, however, is specific to MRI reconstruction and its extension to other inverse problems is not evident.

In this work, we propose and validate a self-supervised method, called \ul{re}covery with a \ul{s}elf-cal\ul{i}brated \ul{de}noiser (ReSiDe), for MRI reconstruction that requires no data beyond the single set of undersampled measurements. The method is broadly applicable to other inverse problems.

\section{method}
\label{sec:method}
\subsection{Compressed sensing and plug-and-play}
We consider the problem of recovering an $N$-pixel image $\bm{x}\in\mathbb{C}^N$ from noisy measurements $\bm{y}\in\mathbb{C}^M$ collected with a known forward operator $\bm{A}\in \mathbb{C}^{M\times N}$, i.e., $\bm{y} = \bm{A}\bm{x} + \bm{\eta}$, where $\bm{\eta}\in\mathbb{C}^M$ is additive white Gaussian noise with variance $\tau^2$. For acceleration rate $R\defn N/M >1$, $\bm{x}$ cannot be uniquely inferred from $\bm{y}$. A typical CS-based formulation to infer $\bm{x}$ from $\bm{y}$ takes the form,
\begin{eqnarray}
\label{eq:cs}
\widehat{\bm{x}} = \underset{\bm{x}}{\argmin}\{\frac{1}{2\tau^2}\|\bm{Ax}-\bm{y}\|_2^2 + \phi(\bm{x})\},
\end{eqnarray}
where $\phi(\bm{x})$ a sparsity-inducing penalty. Typically, $\phi(\bm{x})=\lambda\|\bm{\Psi} \bm{x}\|_1$, where $\bm{\Psi}$ is the wavelet transform, or $\phi(\bm{x})=\lambda \text{TV}(\bm{x})$, where $\text{TV}(\cdot)$ represents total variation. The tuning parameter, $\lambda>0$, controls the regularization strength. Primal-dual splitting (PDS) \cite{ono2017primal} is a popular algorithm to solve Eq.~\ref{eq:cs} and is given in Algorithm~\ref{alg:pds},  with $\bm{f}(\bm{u})=\prox_{\nu\phi}(\bm{u})$, which is the proximal operator defined as
\begin{align}
\prox_{\nu\phi}(\bm{u}) 
&\defn \argmin_{\bm{x}} \left\{\phi(\bm{x}) + \tfrac{1}{2\nu}\|\bm{x}-\bm{u}\|_2^2\right\} 
\label{eq:prox}.
\end{align}
The proximal operator in Eq. \ref{eq:prox} can be interpreted as the maximum a posteriori estimation of $\bm{x}$ from the noisy measurement $\bm{u} = \bm{x}+\bm{w}$, where $\bm{w}$ is additive white Gaussian noise with variance $\nu$, and $\bm{x}$ has the probabilistic prior $p(\bm{x})\propto \exp(-\phi(\bm{x}))$. Besides PDS, the proximal operator also appears in other algorithms commonly used to solve Eq. \ref{eq:cs}, including alternating directions method of multipliers (ADMM). With this denoising viewpoint, Bouman et al. proposed replacing the proximal operator with a call to a powerful denoiser such as BM3D \cite{venkatakrishnan2013plug}. The approach has been called ``plug-and-play'' because essentially any denoiser $\bm{f}(\cdot)$ can be ``plugged into'' ADMM, PDS, or other algorithms used to solve Eq. \ref{eq:cs}.

\begin{algorithm}[h]
  \caption{PnP algorithm}
  \label{alg:pds}
  \begin{algorithmic}[1]
    \REQUIRE $\nu>0$, $\tau$, $\bm{A}$, $\bm{y}$, $\bm{f}$
    \STATE{$\gamma=\frac{\nu}{\tau^2}\|\bm{A}\|_2^{2}$, $\bm{x}_0=\bm{A}^\textsf{H}\bm{y}$, $\bm{z}_0=\bm{Ax}_0-\bm{y}$}
	\FOR{$t=1,2,3,\dots$}
	\STATE{$\bm{u}_t = \bm{x}_{t-1} - \frac{\nu}{\tau^2}\bm{A}^\textsf{H}\bm{z}_{t-1}$}
	\STATE{$\bm{x}_t = \bm{f}(\bm{u}_t)$}
	\label{line:pds_denoise}
	\STATE{$\bm{v}_t = 2\bm{x}_t - \bm{x}_{t-1}$}
	\STATE{$\bm{z}_t=\frac{\gamma}{1+\gamma}\bm{z}_{t-1} + \frac{1}{1+\gamma}(\bm{A}\bm{v}_t - \bm{y})$}
	\ENDFOR
	\RETURN{$\widehat{\bm{x}}\gets\bm{x}_t$}
  \end{algorithmic}
\end{algorithm}

Instead of using generic denoisers (e.g., BM3D or BM4D), it is possible to train application-specific denoisers based on DL and use them in PnP. It has been shown that application-specific denoisers can outperform generic denoisers~\cite{zhang2017beyond}. Also, most generic denoisers are designed only for real-valued 2D images, with limited options available for higher-dimensional or complex-valued images. In contrast, DL denoisers can be trained and implemented for images of any dimension or size.

\subsection{Recovery with a self-calibrated denoiser (ReSiDe)}

Training an application-specific denoiser, $\bm{f}(\cdot)$, in Algorithm \ref{alg:pds} relies on supervised learning, which requires access to clean images. Once trained, the denoiser acts as an implicit prior and can facilitate image recovery when $R>1$. However, clean and high-quality images may not be available for many applications. In response, we propose a PnP-inspired method that iteratively trains the denoiser from the image being recovered. The resulting method, called ReSiDe, is implemented using a modified PDS algorithm (Algorithm \ref{alg:pds-selfcali}).

\begin{algorithm}[h]
  \caption{ReSiDe algorithm}
  \label{alg:pds-selfcali}
  \begin{algorithmic}[1]
    \REQUIRE $\nu>0$, $\tau$, $\bm{A}$, $\bm{y}$, $\bm{f}$
    \STATE{$\gamma=\frac{\nu}{\tau^2}\|\bm{A}\|_2^{2}$, $\bm{x}_0=\bm{A}^\textsf{H}\bm{y}$, $\bm{z}_0=\bm{Ax}_0-\bm{y}$}
	\FOR{$t=1,2,3,\dots$}
	\STATE{$\tilde{\bm{x}}_{t-1} = \bm{x}_{t-1} + \mathcal{N}(\bm{0}, \sigma_t^2 \bm{I})$}
	\STATE{$\bm{\theta}_t = \operatorname{argmin}_{\bm{\theta}} \sum_{i=1}^P \mathcal{L}(\bm{f}(\mathcal{I}[\tilde{\bm{x}}_{t-1}]_i; \bm{\theta}), \mathcal{I}[\bm{x}_{t-1}]_i)$}
	\STATE{$\bm{u}_t = \bm{x}_{t-1} - \frac{\nu}{\tau^2}\bm{A}^\textsf{H}\bm{z}_{t-1}$}
	\STATE{$\bm{x}_t = \bm{f}(\bm{u}_t; \bm{\theta}_t)$}
	\label{line:pds_denoise}
	\STATE{$\bm{v}_t = 2\bm{x}_t - \bm{x}_{t-1}$}
	\STATE{$\bm{z}_t=\frac{\gamma}{1+\gamma}\bm{z}_{t-1} + \frac{1}{1+\gamma}(\bm{A}\bm{v}_t - \bm{y})$}
	\ENDFOR
	\RETURN{$\widehat{\bm{x}}\gets\bm{x}_t$}
  \end{algorithmic}
\end{algorithm}

In Algorithm \ref{alg:pds-selfcali}, the current estimate, $\bm{x}_{t-1}$, is contaminated with complex-valued, zero-mean white Gaussian noise of variance $\sigma_t^2$ to generate a noisy image, $\tilde{\bm{x}}_{t-1}$ (Line 3). Then, we train a network by feeding it $P\geq1$ pairs of noisy-clean patches as input-output (Line 4). The noisy patches are extracted from $\tilde{\bm{x}}_{t-1}$, while the clean patches are extracted from ${\bm{x}}_{t-1}$. The locations of the patches within $\bm{x}_{t-1}$ or $\tilde{\bm{x}}_{t-1}$ are selected randomly. The patch extraction is performed by the operator $\mathcal{I}$, with $\mathcal{I}[\cdot]_i$ representing the $i^\text{th}$ patch. The network is parameterized by $\bm{\theta}$, with $\bm{\theta}_t$ representing the network parameters at the $t^\text{th}$ iteration. In each iteration, the network is trained de novo from a random initialization of the network parameters. Finally, we use the trained network to denoise $\bm{u}_t$ (Line 6). Here, the operator $\mathcal{L}(\cdot)$ defines the loss function used in training. We chose $\mathcal{L}(\bm{f}(\mathcal{I}[\tilde{\bm{x}}_{t-1}]_i; \bm{\theta}), \mathcal{I}[\bm{x}_{t-1}]_i)=\|\bm{f}(\mathcal{I}[\tilde{\bm{x}}_{t-1}]_i; \bm{\theta}) - \mathcal{I}[\bm{x}_{t-1}]_i \|_2^2$.

\section{EXPERIMENT AND RESULTS}
\label{sec:experiment and results}
\subsection{Experimental settings}
\label{ssec:exp}
For validation, two brain datasets are used, i.e., a $320 \times 320$ T1-weighted and a $384 \times 384$ T2-weighted brain MRI. The multi-coil k-space data were downloaded from fastMRI (\url{https://fastmri.org/}) and converted to complex-valued single-coil images using ESPIRiT-based sensitivity maps \cite{uecker2014espirit}. The k-space corresponding to the single-coil image was undersampled at $R=1.8$ using two different Cartesian sampling masks: variable density pseudo-random sampling with undersampling in only one direction (M1), shown in Figures \ref{fig:T1_gro} and \ref{fig:T2_gro}, and random sampling with undersampling in both directions (M2), shown in Figures \ref{fig:T1_random} and \ref{fig:T2_random}. For both sampling patterns, 32 central lines were fully sampled.

The denoiser architecture is similar to one we have previously reported for supervised training~\cite{ahmad2020plug}. In summary, the network architecture is inspired by DnCNN \cite{zhang2017beyond}, with five convolutional layers, four ReLu layers in between, and a skip connection. The first four convolutional layers have 64 $3\times3$ kernels, and the last convolutional layer has two $3\times3$ kernels. We split real and imaginary parts into two channels.

In each iteration, we selected $P=144$ patches of size $64\times64$.  We decreased the noise standard deviation, $\sigma_t$, after every ten iterations such that the training signal-to-noise ratio, $\text{SNR}_t = 20\log_{10}(\|\bm{x}_t\|_2/(\sqrt{2N}\sigma_t))$, increased from 10 dB to 40 dB in the increments of 5~dB after every ten iterations, i.e., $\text{SNR}_t=10 + 5 \left \lfloor{(t-1)/10}\right \rfloor$, where $\left \lfloor{k}\right \rfloor$ rounds $k$ to the nearest integer less than or equal to $k$. The training was performed on a workstation equipped with a single NVIDIA GPU (GeForce RTX 2080 Ti). We set the minibatch size to 16 and used the Adam optimizer with a learning rate of $1 \times 10^{-3}$. The number of epochs was fixed at 100, and the algorithm was executed for 70 iterations.

To compare ReSiDe with other unsupervised methods, we also recovered the images using a CS method, called L1-wavelet, and PnP-BM3D. The L1-wavelet reconstruction was performed using the BART toolbox (\url{https://mrirecon.github.io/bart/}), while PnP-BM3D was performed using Algorithm \ref{alg:pds}, with $\bm{f}(\cdot)$ representing a call to the BM3D denoiser, which was applied separately to real and imaginary parts of the image. BM3D also takes noise variance as an input; this parameter was manually tuned and left fixed for all experiments. The performance evaluation was based on NMSE defined as $20\log_{10}(\|\bm{x}-\widehat{\bm{x}}\|_2/\|\bm{x}\|_2)$. 




\subsection{Results}
\label{sec:result}
\begin{figure}[h!]
    \centering
    \includegraphics[width = \columnwidth]{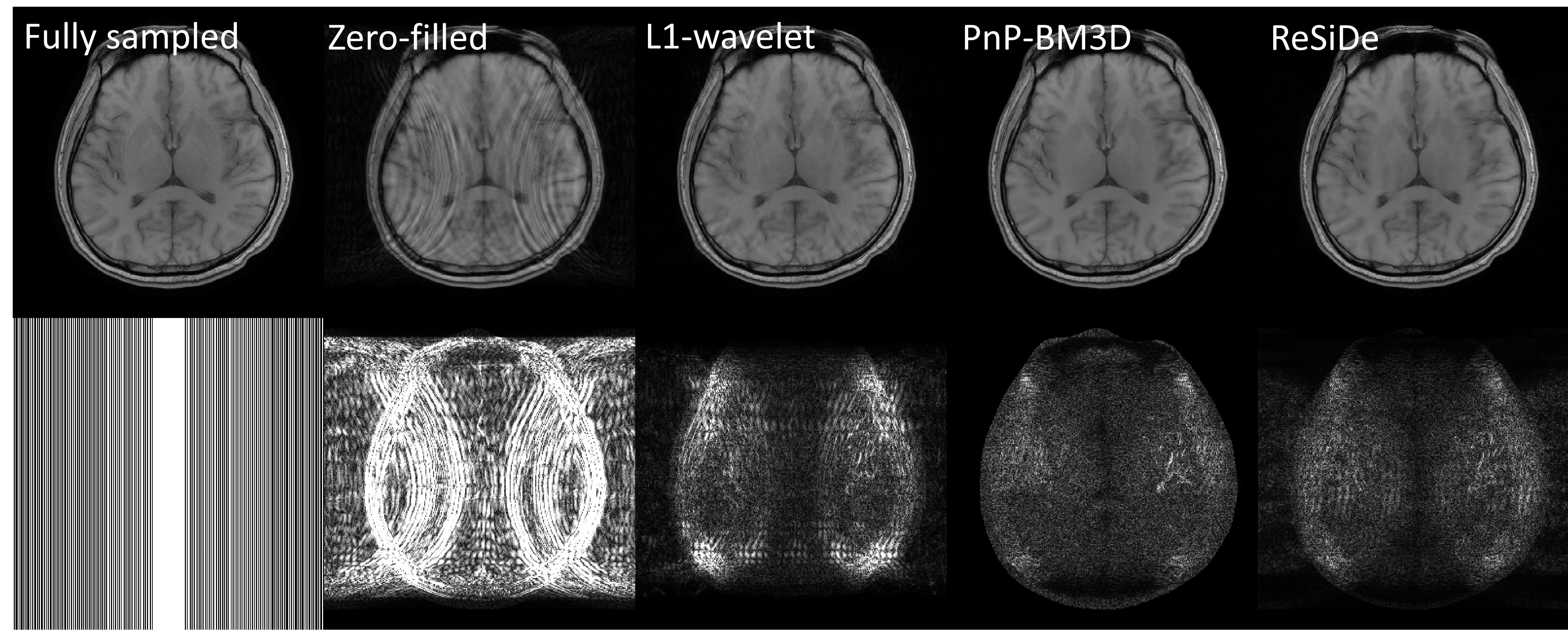}
    \caption{T1-weighted brain image reconstruction using the structured sampling (M1). The second row shows the absolution error map after 1.5-fold amplification.}
    \label{fig:T1_gro}
\end{figure}

\begin{figure}[h!]
    \centering
    \includegraphics[width = \columnwidth]{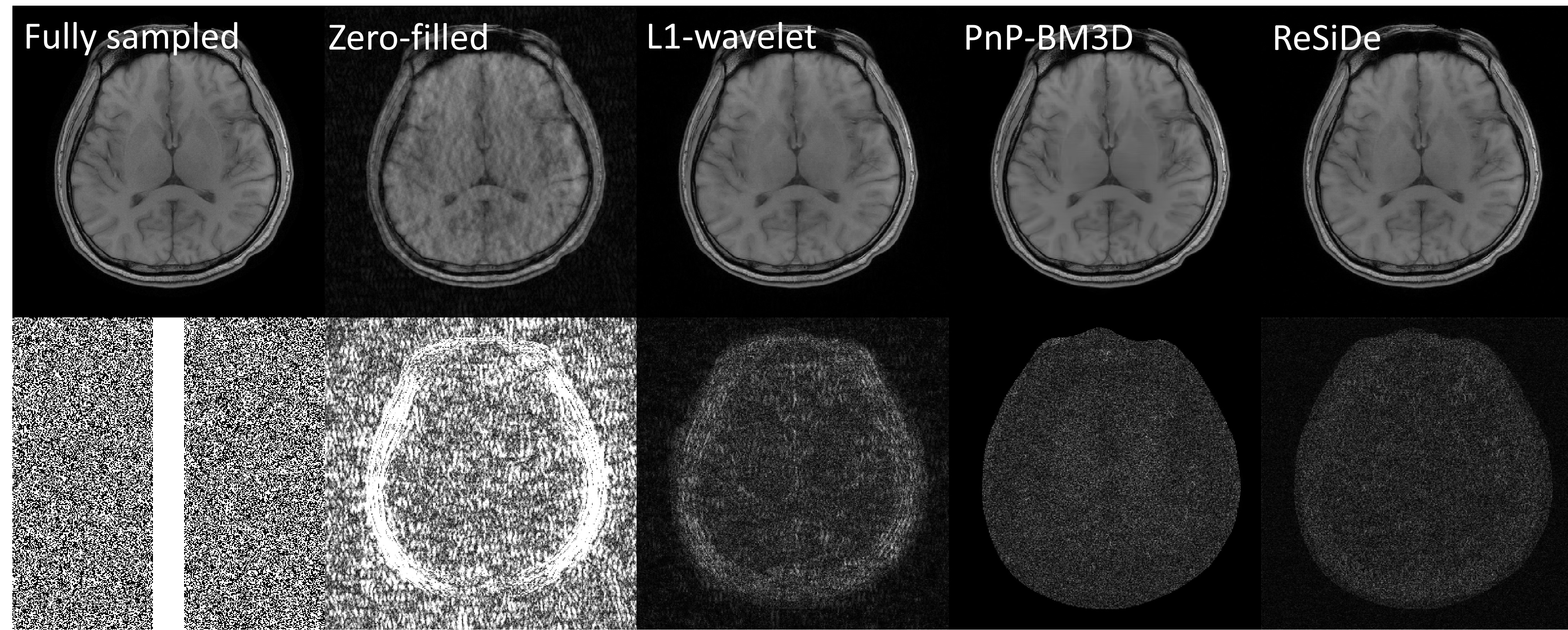}
    \caption{Same as Figure \ref{fig:T1_gro} but with the random sampling (M2).}
    \label{fig:T1_random}
\end{figure}

\begin{figure}[h!]
    \centering
    \includegraphics[width = \columnwidth]{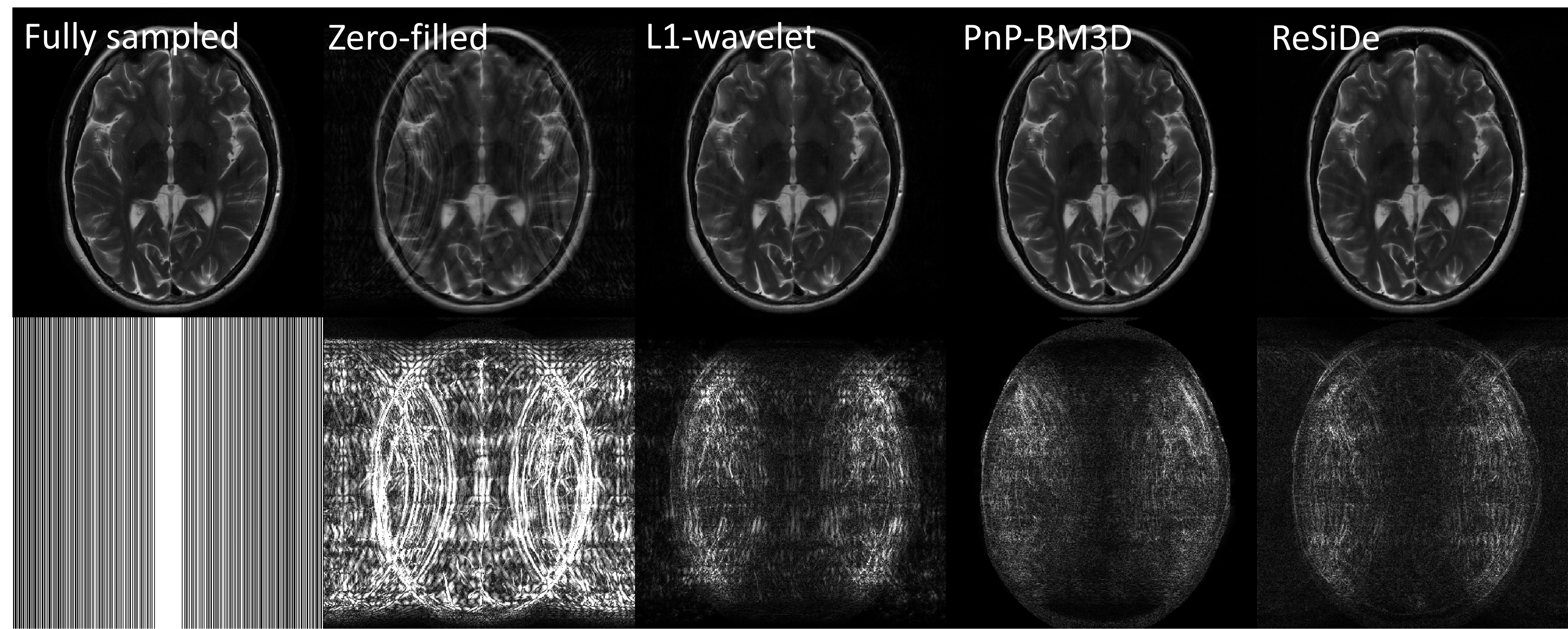}
    \caption{T2-weighted brain image reconstruction using the structured sampling (M1). The second row shows the absolution error map after 1.5-fold amplification.}
    \label{fig:T2_gro}
\end{figure}

\begin{figure}[h!]
    \centering
    \includegraphics[width = \columnwidth]{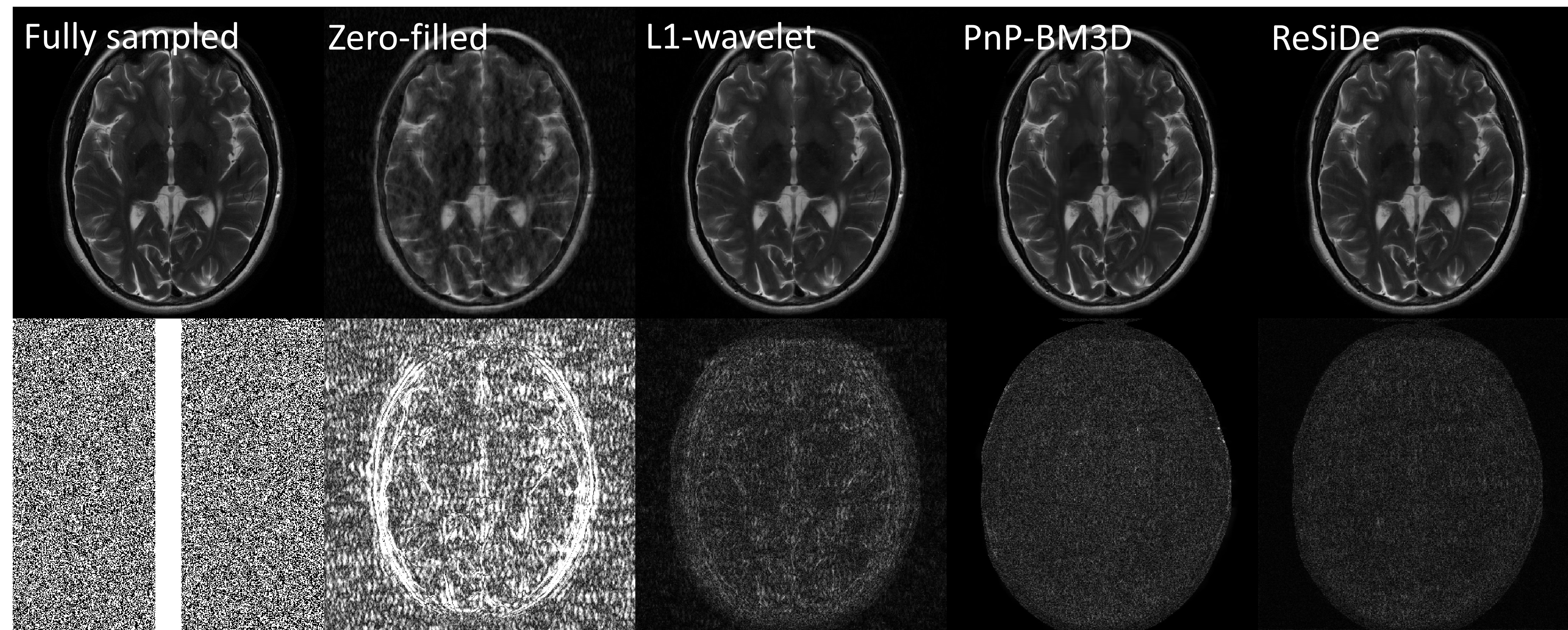}
    \caption{Same as Figure \ref{fig:T2_gro} but with the random sampling (M2).}
    \label{fig:T2_random}
\end{figure}

\begin{figure}[h!]
    \centering
    \includegraphics[width = 0.95\columnwidth]{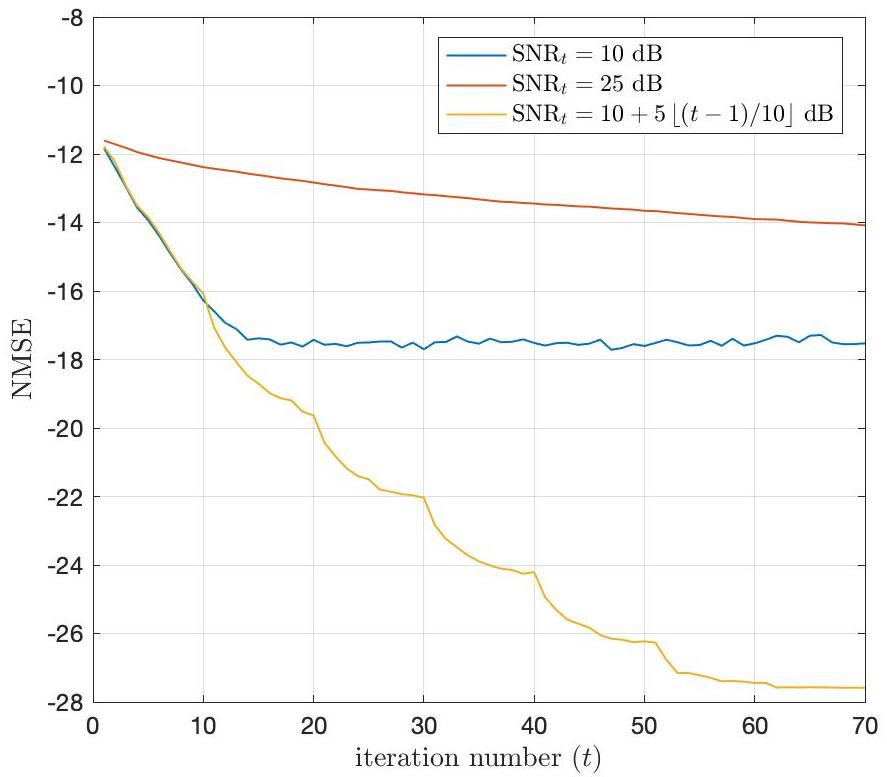}
    \caption{Change in NMSE of ReSiDe with the iteration number. The results are shown for the T2-weighted image with random sampling pattern (M2) and three different $\text{SNR}_t$ settings. Similar behavior was observed for other three combinations of sampling and image contrast.}
    \label{fig:nmse_curve}
\end{figure}

Figures \ref{fig:T1_gro}--\ref{fig:T2_random} show the reconstructions for T1 and T2 images from two different sampling patterns, M1 and M2. The first row includes the true image from fully sampled k-space, zero-filled reconstruction, and images from L1-wavelet, PnP-BM3D, and ReSiDe reconstructions. The second row shows the sampling pattern along with the corresponding error maps. Table 1 summarizes NMSE values from different reconstruction methods. Overall, ReSiDe outperforms L1-wavelet and PnP-BM3D in all four cases, with an average advantage of 3.06~dB over L1-wavelet and 0.91~dB over PnP-BM3D. All three reconstruction methods perform better for the sampling pattern that is less structured (M2). Figure \ref{fig:nmse_curve} shows the convergence curve of ReSiDe under three different $\text{SNR}_t$ settings, i.e., a fixed $\text{SNR}_t$ of 10 dB, a fixed $\text{SNR}_t$ of 25 dB, and $\text{SNR}_t$ that progressively increases over iterations.


\begin{table}[t]
\smallskip
\centering
\begin{adjustbox}{width=0.45\textwidth}
\begin{tabular}{|l|l|l|l|l|l|}
\hline
\bf{Contrast}  & \bf{Sampling} &\bf{Zero-filled} & \bf{L1-wavelet} & \bf{PnP-BM3D} & \bf{ReSiDe} \\ \hline \hline
$T1$  & M1    & $-13.50$ & $-24.31$ & $-27.37$ & $\bf{-27.59}$\\ \hline
$T1$  & M2    & $-12.97$ & $-26.97$ & $-28.82$ & $\bf{-29.96}$\\ \hline
$T2$  & M1    & $-11.86$ & $-21.67$ & $-23.60$ & $\bf{-24.08}$\\ \hline
$T2$  & M2    & $-11.52$ & $-24.04$ & $-25.81$ & $\bf{-27.59}$\\ \hline
\end{tabular}
\end{adjustbox}
\caption{NMSE (dB) for the T1 and T2-weighted brain MRI data undersampled with M1 and M2 sampling patterns.}
\label{tab:rsnr}
\end{table}

\section{Discussion}
\label{sec:discussion}
MRI reconstruction methods have evolved rapidly in the last two decades, with the DL-based methods taking over CS-based methods in the latest, ongoing round. Most DL methods used for MRI recovery require access to a large corpus of high-quality training data in the form of raw k-space data or images. There are several MRI applications, e.g., contrast-based dynamic applications, where obtaining such training data is challenging or infeasible. To facilitate such applications, we propose a self-calibrated scheme called ReSiDe. In contrast to the traditional PnP approach, where an off-the-shelf or an application-specific trained denoiser is used, ReSiDe iteratively trains the denoiser from the image being recovered, obviating the need for training data while enabling image-specific training of the denoiser. For validation, ReSiDe is applied to single-coil MRI data with two different contrasts and two different sampling patterns.


We observed ReSiDe to converge for a wide range of $\sigma_t$ values. However, the speed of convergence and the final NMSE were greatly impacted by $\sigma_t$. Selecting a fixed, low training $\text{SNR}_t$ (e.g., 10 dB) improved the convergence speed but led to lower final NMSE and generated overly smooth images. In contrast, selecting a fixed, high training $\text{SNR}_t$ (e.g., 25 dB) improved the final NMSE but offered a very slow convergence. To improve both the convergence speed and the final NMSE, we adopted a scheme where the starting $\text{SNR}_t$ of 10~dB is increased by 5~dB after every ten iterations (Figure \ref{fig:nmse_curve}). 

A major limitation of ReSiDe is the reconstruction time, with each iteration taking approximately 30 minutes on a workstation with a single GPU. ReSiDe shares this limitation with other training-free DL methods such as deep image prior~\cite{ulyanov2018deep}. The computation speed of ReSiDe can be potentially improved by training the denoiser after every $k^\text{th}$ iteration, with $k>1$, simplifying the denoiser architecture, retaining network weights from one iteration to the next, and transitioning $\sigma_t$ based on the relative change in the image, i.e., $\|\bm{x}_t - \bm{x}_{t-1}\|_2 / \|\bm{x}_t \|_2$. We also believe that the performance of ReSiDe can be further improved by optimizing the denoiser architecture and employing a different loss function. The architecture used in this work is identical to the one we previously used to train an application-specific denoiser from a large number of training patches. Optimizing the denoiser architecture and evaluating various loss functions as well as application to multi-coil MRI will be considered in future studies. 

In this work, we have presented a very specific method to train the denoiser, i.e., we add noise to $\bm{x}_{t-1}$ and train the denoiser to remove this noise. More recently, several ``blind denoising'' methods (e.g., Noise2Void~\cite{krull2019noise2void} and Noise2Self~\cite{batson2019noise2self}) have been proposed that also train a network (denoiser) from a single noisy image. In general, these methods employ cross-validation to train a network to estimate one subset of pixels from another subset. However, to the best of our knowledge, these methods have not been utilized to solve broader inverse problems where $\bm{A}\neq \bm{I}$. One exception is the method termed as Noise2Inverse, where a self-calibrated denoiser (Noise2Self) is used to recover CT images \cite{hendriksen2020noise2inverse}. However, in Noise2Inverse, the denoising step is applied to the images recovered with filtered backprojection and is not integrated into the reconstruction process. Nonetheless, we believe other self-calibrated denoisers, including Noise2Void and Noise2Self, can also be used within ReSiDe and may offer faster convergence or lower NMSE.

\section{Conclusion}
\label{sec:conclusion}
A self-calibrated method is proposed for MRI reconstruction. The proposed method, called ReSiDe, outperforms other training-free methods, namely L1-wavelet and PnP-BM3D, in terms of image quality. Although ReSiDe is described and applied to single-coil MRI in this work, it is equally applicable to multi-coil MRI and other inverse problems.



\bibliographystyle{IEEEbib}
\bibliography{root.bib}

\end{document}